\documentclass[fleqn,usenatbib]{mnras}

\usepackage{newtxtext,newtxmath}

\usepackage[T1]{fontenc}

\DeclareRobustCommand{\VAN}[3]{#2}
\let\VANthebibliography\thebibliography
\def\thebibliography{\DeclareRobustCommand{\VAN}[3]{##3}\VANthebibliography}

\usepackage[usenames,dvipsnames,svgnames,table]{xcolor}

\usepackage{graphicx}	\usepackage{amsmath}	\usepackage{amssymb}

\title[MAD Sgr A* flares]{Sgr A* near-infrared flares from reconnection events in a magnetically arrested disc}

\author[Dexter, Tchekhovskoy, Jim\'{e}nez-Rosales, Ressler, et al.]{
J. Dexter,$^{1,2}$\thanks{jason.dexter@colorado.edu}
A. Tchekhovskoy,$^{3}$
A. Jim\'{e}nez-Rosales,$^{2}$
S. M. Ressler,$^{4}$
M. Baub\"{o}ck,$^{2}$
\newauthor
Y. Dallilar,$^{2}$ P. T. de Zeeuw,$^{2,5}$ F. Eisenhauer,$^{2}$ S. von Fellenberg,$^{2}$ F. Gao,$^{2}$ R. Genzel,$^{2,6}$\newauthor
S. Gillessen,$^{2}$ M. Habibi,$^{2}$ T. Ott,$^{2}$ J. Stadler,$^{2}$ O. Straub,$^{2}$ F. Widmann$^{2}$
\\
$^{1}$JILA and Department of Astrophysical and Planetary Sciences, University of Colorado, Boulder, CO 80309, USA\\
$^{2}$Max Planck Institute for Extraterrestrial Physics (MPE), Giessenbachstr. 1, 85748 Garching, Germany\\
$^{3}$Center for Interdisciplinary Exploration \& Research in Astrophysics (CIERA), Physics \& Astronomy, Northwestern University,\\ Evanston, IL 60202, USA\\
$^{4}$Kavli Institute for Theoretical Physics, University of California Santa Barbara, Kohn Hall, Santa Barbara, CA 93107, USA\\
$^{5}$Sterrewacht Leiden, Leiden University, Postbus 9513, 2300 RA Leiden, The Netherlands\\
$^{6}$Departments of Physics and Astronomy, Le Conte Hall, University of California, Berkeley, CA 94720, USA
}

\date{Accepted XXX. Received YYY; in original form ZZZ}

\pubyear{2019}

\begin{document}
\label{firstpage}
\pagerange{\pageref{firstpage}--\pageref{lastpage}}
\maketitle

\begin{abstract}
Large-amplitude Sgr A* near-infrared flares result from energy injection into electrons near the black hole event horizon. Astrometry data show continuous rotation of the emission region during bright flares, and corresponding rotation of the linear polarization angle. One broad class of physical flare models invokes magnetic reconnection. Here we show that such a scenario can arise in a general relativistic magnetohydrodynamic simulation of a magnetically arrested disc. Saturation of magnetic flux triggers eruption events, where magnetically dominated plasma is expelled from near the horizon and forms a rotating, spiral structure. Dissipation occurs via reconnection at the interface of the magnetically dominated plasma and surrounding fluid. This dissipation is associated with large increases in near-infrared emission in models of Sgr A*, with durations and amplitudes consistent with the observed flares. Such events occur at roughly the timescale to re-accumulate the magnetic flux from the inner accretion disc, $\simeq 10$h for Sgr~A*. We study near-infrared observables from one sample event to show that the emission morphology tracks the boundary of the magnetically dominated region. As the region rotates around the black hole, the near-infrared centroid and linear polarization angle both undergo continuous rotation, similar to the behavior seen in Sgr~A* flares. 
\end{abstract}

\begin{keywords}
accretion, accretion discs --- black hole physics --- Galaxy: centre --- MHD --- polarization --- radiative transfer
\end{keywords}

\begin{figure*}
\includegraphics[width=0.95\textwidth]{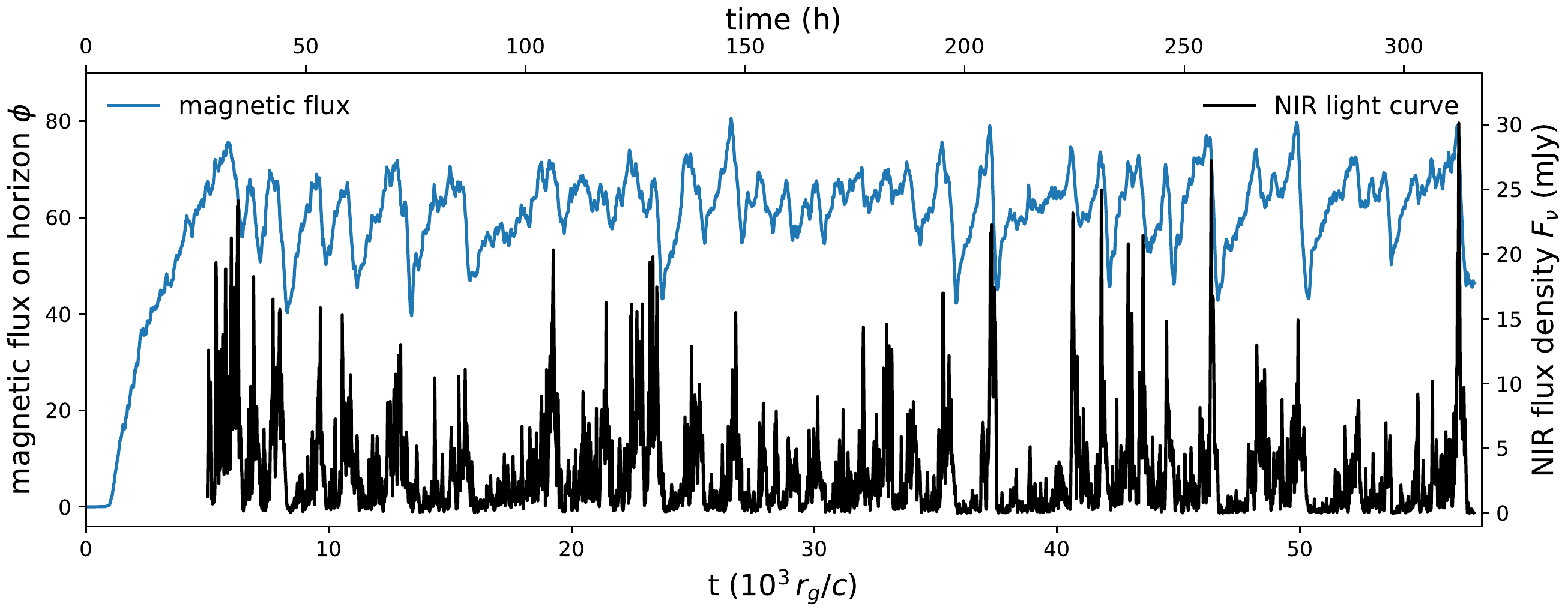}
\caption{\label{fig:sim_plots_phibh_lcurve}Magnetic flux accumulated on the event horizon as a function of time (blue) and the near-infrared light curve (black) for our fiducial MAD simulation. The magnetic flux shows cycles of gradual ramp up and rapid dissipation. The dissipation events are due to magnetic flux eruptions from near the black hole, and are associated with large-amplitude near-infrared variability. Such ``flares'' occur every $\simeq 10$h for Sgr~A*. The NIR light curve is plotted as $F_\nu / \langle\dot{M}\rangle^2$ with $\langle\dot{M}\rangle$ the temporally smoothed accretion rate. This keeps the median flux density roughly constant as the mass reservoir drains.}
\end{figure*}

\section{Introduction}

The Galactic center massive black hole, Sgr A*, shows variable near-infrared (NIR) emission with factor $\sim 10$ increases \citep{genzel2003,ghez2004} over its median value \citep{schoedel2011,doddseden2011,witzel2018}. The so-called ``flares'' are often accompanied by simultaneous events in the X-ray \citep{baganoff2001,eckart2008sim,marrone2008}. The lack of a clear submillimeter counterpart (submm, electron energy $\gamma \sim 10-100$) implies that the flares are due to the acceleration of electrons to energies $\gamma \gtrsim 10^3$ \citep{markoff2001}. Linear polarization fractions of $\simeq 10$--$40\%$ indicate that the NIR emission is due to synchrotron radiation  \citep{eckart2006,trippe2007,eckart2008pol}. Despite nearly two decades of study, their physical origin remains uncertain.

Recently, NIR long baseline interferometry with the VLT Interferometer instrument GRAVITY \citep{gravity2017} showed a continuous rotation of the NIR centroid during three bright flares with apparent periods of $\simeq 30$--$60$ minutes \citep{gravity2018flare}. The observed clockwise motion on sky is consistent with models of a compact orbiting emission region \citep{broderickloeb2005,broderick2006} at a radius of $r \simeq 6$--$10 r_g$, where $r_g = GM/c^2$ is the gravitational radius \citep[][but see \citealt{matsumoto2020} and \citealt{ball2020}]{bauboeck2020}. Simultaneous rotation of the linear polarization angle with a comparable period is consistent with the same scenario, as long as there is a significant poloidal magnetic field component in the emission region \citep[][Gravity Collaboration et al., 2020, submitted]{gravity2018flare}. The circular pattern of the centroid motion on sky and lack of a strong Doppler beaming signature in the flares disfavors inclination angles close to edge-on \citep[$i \lesssim 130^\circ$,][]{bauboeck2020}.

Radiative models based on general relativistic magnetohydrodynamic (GRMHD) simulations of Sgr A* accretion are consistent with the source spectrum, image sizes, and image-integrated polarization properties \citep{moscibrodzka2009,moscibrodzka2014,dexter2009,dexter2010,shcherbakov2012,chan2015image,ressler2017,chael2018,anantua2020}. In some cases, the models produce sufficiently hot electrons to match the observed NIR luminosity \citep[e.g.,][]{dexter2013,chan2015var,ressler2017}. The NIR emission region is usually found to be concentrated close to the black hole event horizon \citep[e.g.,][]{dolence2009,ressler2017,petersen2020}. 

Here we consider a scenario for Sgr A* flares as the result of stochastic, repeating, large-scale magnetic reconnection events occurring in GRMHD models of magnetically arrested discs \citep[MADs,][]{bisnovatyi1974,igumenshchev2003,narayan2003,tchekhovskoy2011,mckinney2012}. We study one GRMHD model that we found to be broadly consistent with observations of Sgr~A* \citep{dexter2020harmpi}. We show that NIR flares occur every $\simeq 10$h in the same models as the result of magnetic eruptions originating close to the black hole (\autoref{sec:madflares}). The flares show continuous rotation of the astrometric centroid as a result of rotating spiral structure in the emission region (\autoref{sec:niremission}). We find a corresponding rotation of the polarization angle due to the strong poloidal fields near the black holes. We discuss the limitations of the current model and implications for our understanding of accretion onto Sgr~A* (\autoref{sec:discussion}). We note that \citet{porth2020} have also carried out a study of these events and their possible connection to Sgr A* flares, including calculations of their dynamics, energetics, magnetic field configuration, and dependence on black hole spin.

\begin{figure*}
\includegraphics[width=0.9\textwidth]{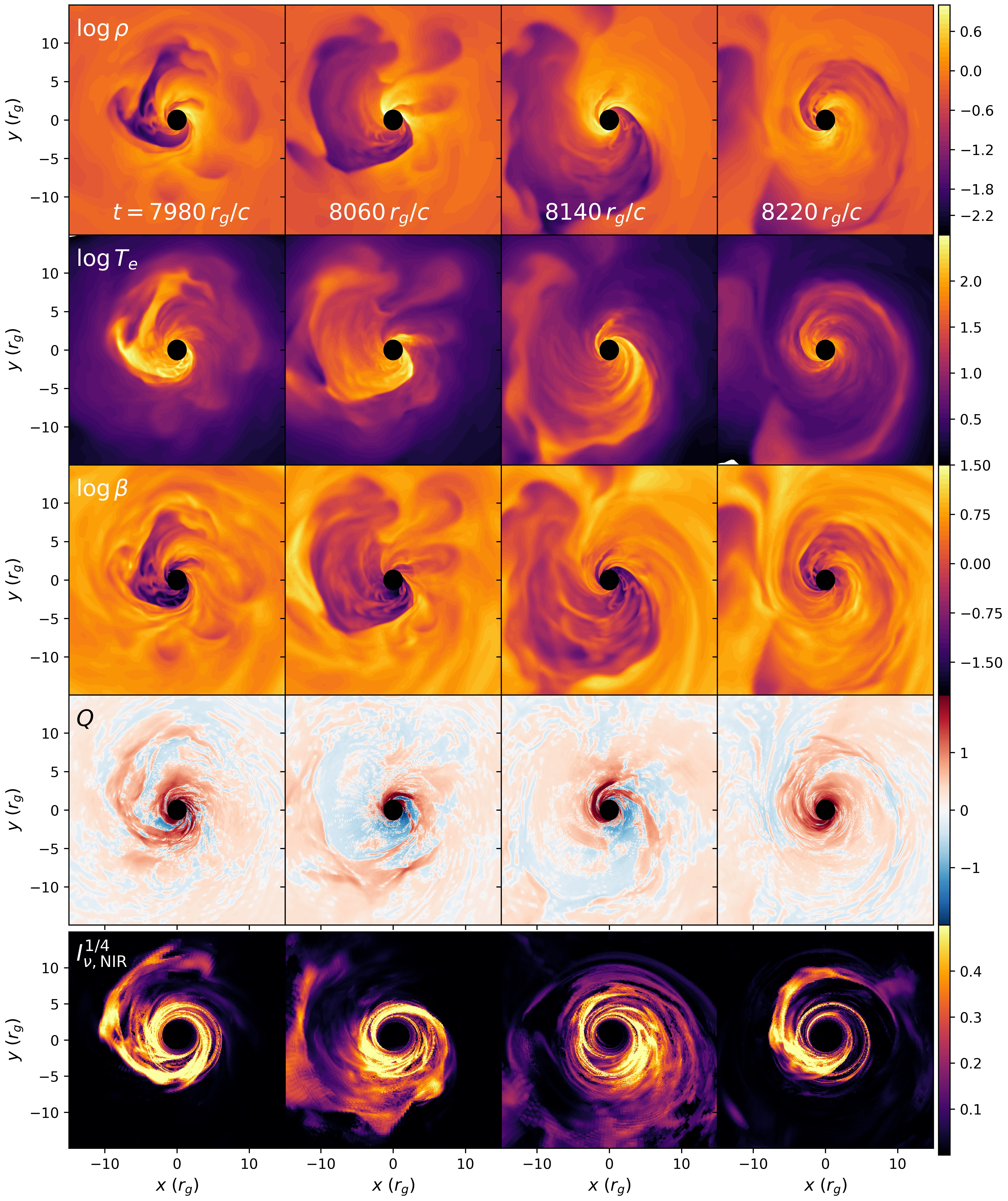}
\caption{\label{fig:sim_plots_temaps}Vertically averaged maps of simulation variables of density $\rho$, dimensionless electron temperature $k T_e / m_e c^2$, plasma $\beta$, and fluid heating rate $Q$ at four snapshots during the flare period near $8 \times 10^3 \, r_g/c$. The density and heating rate are in code units. Ray traced NIR images are shown in the bottom row, scaled as intensity $I_\nu^{1/4}$ and normalized to the maximum pixel brightness of each image. Strong heating occurs continuously at the interface of the magnetically dominated, erupting regions and surrounding fluid. The rotating, spiral morphology of the flaring region matches that of the eruption, and the radiation originates at the interface where strong heating occurs via magnetic reconnection.}
\end{figure*}

\begin{figure*}
\begin{tabular}{ccc}
\includegraphics[width=0.45\textwidth]{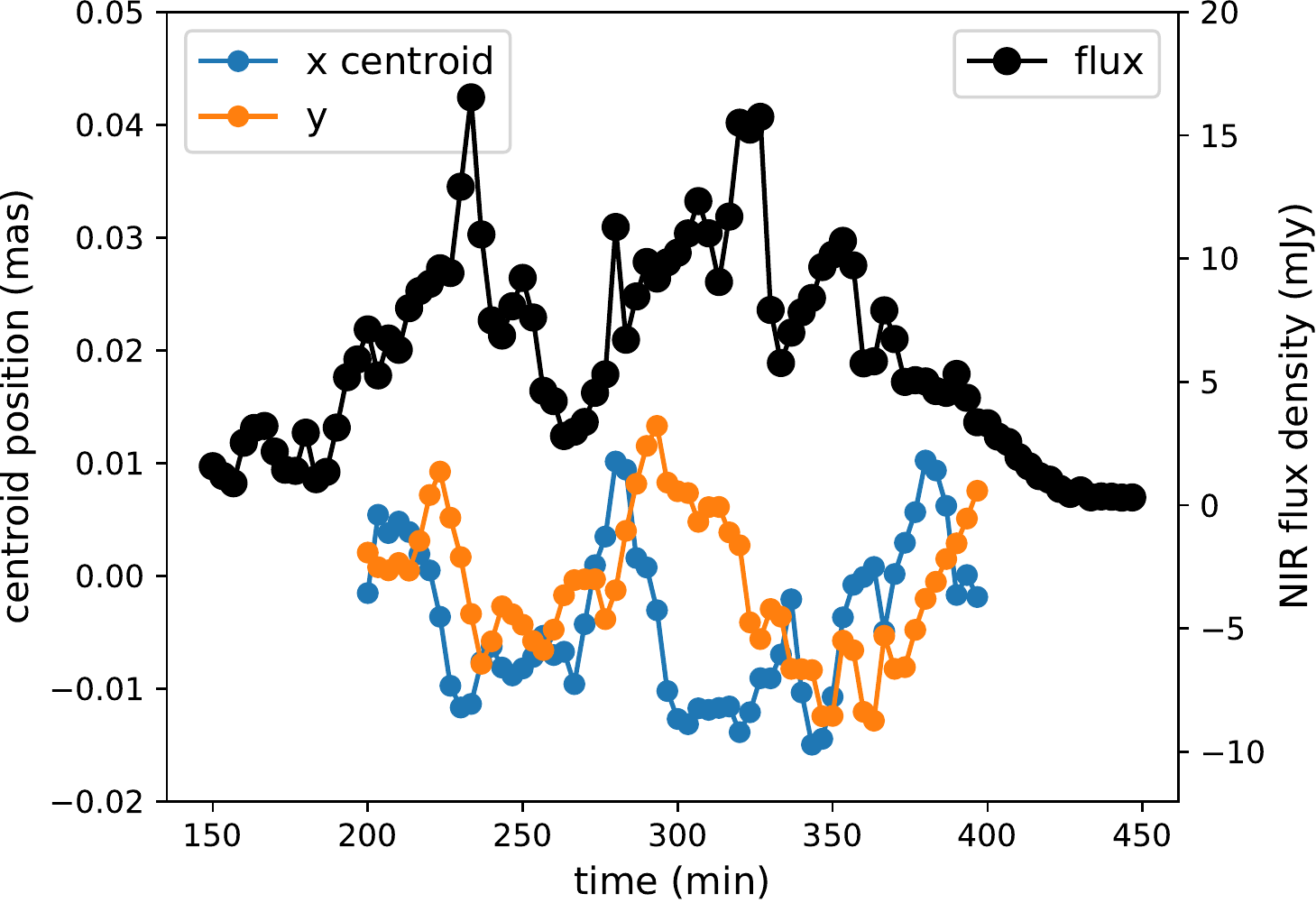} &
\includegraphics[width=0.4\textwidth]{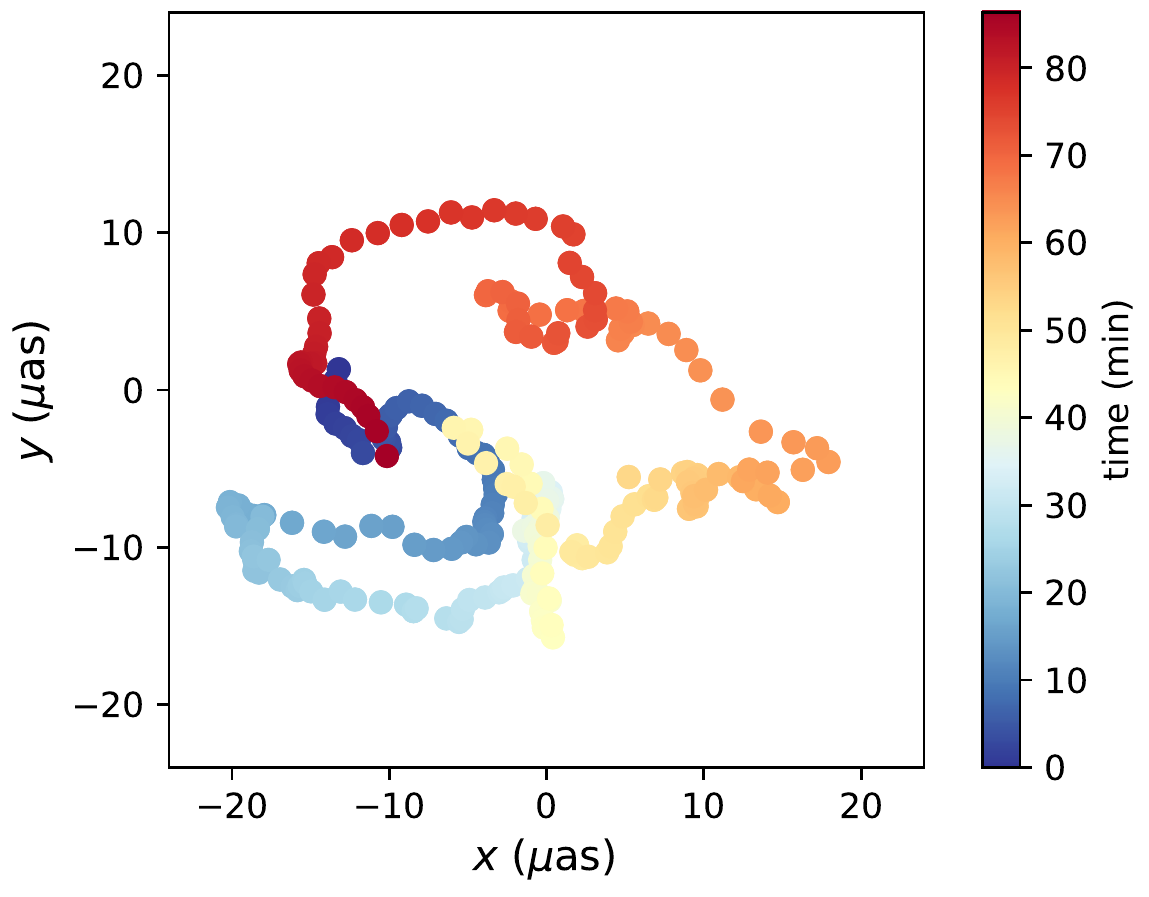}
\end{tabular}
\caption{\label{fig:flux_astro}Left: Total NIR flux density (black) and $x$ and $y$ image centroid positions (blue and orange) as a function of time during the MAD flux eruption event near $t = 8 \times 10^3 r_g/c$ in our fiducial simulation. The rapid, large amplitude variability is accompanied by rotation in the centroid, particularly during the end of the flare from $\simeq 320$--$400$ minutes. Right: The NIR $x$ and $y$ total intensity centroid positions are shown color-coded in time from blue to yellow to red over the period of $320-400$ minutes of the flare from the left panel. The NIR centroid rotates continuously with an astrometric period of $\simeq 80$ minutes.}
\end{figure*}

\begin{figure*}
\begin{tabular}{ccc}
\includegraphics[width=0.4\textwidth]{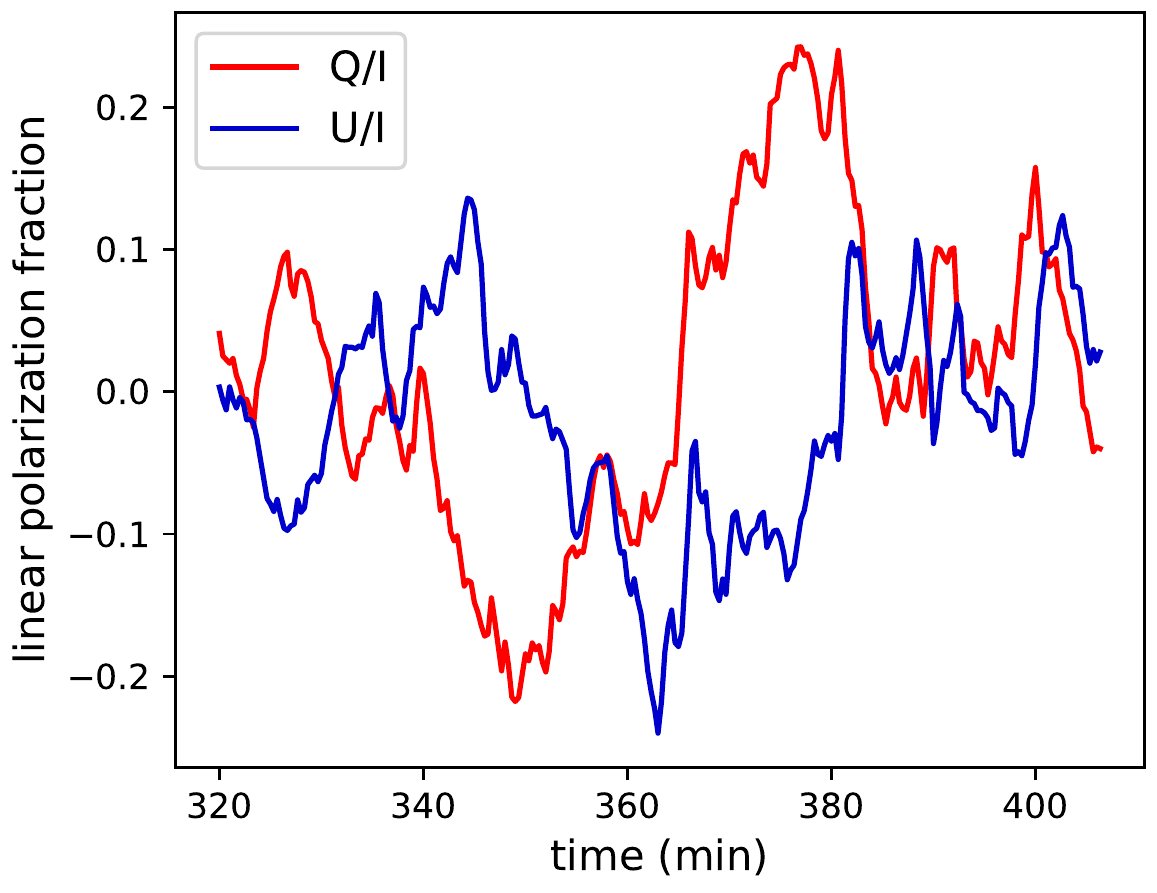}
&
\includegraphics[width=0.4\textwidth]{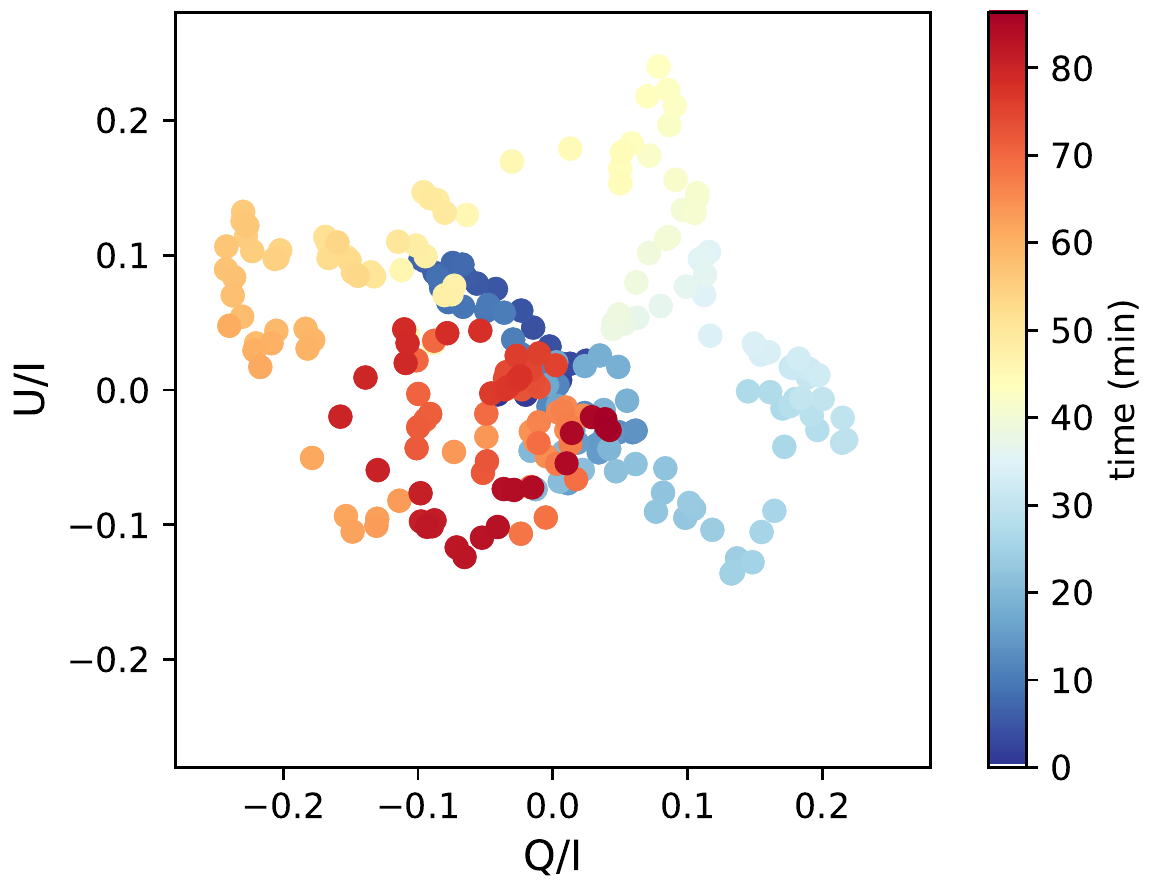}
\end{tabular}
\caption{\label{fig:qu}Fractional linear polarization in Stokes $Q/I$ and $U/I$ over the flare time period of $320-400$ min plotted vs. time (left) and in the $Q/U$ plane color-coded in time in the same sense as the right panel of \autoref{fig:flux_astro} (right). The polarization angle rotates with a comparable period as the astrometric motion, as seen in the $Q/I$ and $U/I$ oscillations and circular trajectory in the  $Q/I$, $U/I$ plane.}
\end{figure*}

\section{Flux eruptions in numerical models of magnetically arrested discs}
\label{sec:madflares}

Here we study one fiducial, long duration, MAD, 3D GRMHD simulation described in \citet{dexter2020harmpi} and run with the \texttt{harmpi}\footnote{\url{https://github.com/atchekho/harmpi}} code \citep{harmpiascl}. Using a resolution of $320 \times 256 \times 160$ grid cells in the $r-$, $\theta-$, and $\varphi-$directions, respectively, the simulation was run for a time of $6\times10^4 r_g/c$ after being initialized from a Fishbone-Moncrief torus with inner radius $r_{\rm in}=12 r_g$, pressure maximum radius $r_{\rm max}=25 r_g$, and black hole spin parameter of $a=0.9375$. A single poloidal loop of magnetic field particularly designed to supply a large amount of magnetic flux was added inside the torus such that $\max(p_g)/\max(p_B) = 100$, where $p_g$ and $p_B$ are the gas and magnetic pressures (see, e.g., \citealt{tchekhovskoy2011}). By the end of the run, inflow equilibrium was established out to $r \simeq 70 \, r_g$, somewhat lower than the $\simeq 110 \, r_g$ found in the long duration MAD simulation of \citet{narayan2012}. The simulation included a scheme for self-consistently evolving four separate electron internal energy densities along with that of the single MHD fluid \citep{ressler2015}. We use an ideal fluid equation of state with adiabatic index of $5/3$ for the fluid, and $4/3$ for the electrons. Following \citet{ressler2017} and \citet{chael2018}, we implemented four different electron internal energies using electron heating models based on both turbulent and magnetic reconnection kinetics calculations. Here we focus on the magnetic reconnection model of \citet[][electron heating fraction of $q_e = 1/4 - 1/2$, parameterized using their equation 3]{werner2018}, which can produce both the median NIR flux density and large-amplitude flaring behavior. The results presented here are consistent across all electron heating models.

The blue line in \autoref{fig:sim_plots_phibh_lcurve} shows the dimensionless magnetic flux accumulated on the horizon as a function of time, $\phi = \sqrt{4\pi} \Phi_{\rm BH} / \sqrt{\dot{M}}$, where $\Phi_{\rm BH}$ is the magnetic flux and $\dot{M}>0$ is the accretion rate which we smooth over timescales of $1000 \, r_g/c$. Magnetic flux is advected inwards with the flow and rapidly builds up on the event horizon, saturating in a MAD state where $\phi \simeq 50-60$  \citep[e.g.,][]{tchekhovskoy2011,mckinney2012,2015ASSL..414...45T}. The normalized magnetic flux undergoes cycles of gradual build up and rapid dissipation. The dissipation is due to stochastic magnetic flux  ``eruption'' events from near the black hole \citep{igumenshchev2008}. 

The flux eruptions occur after the time required to replenish the magnetic flux lost by the black hole during the previous event. We can estimate the recurrence timescale for the largest eruptions, in which the black hole loses about half of its magnetic flux, as the time it takes for the accreting gas (with frozen-in magnetic flux) to reach the black hole from $r = 20 \, r_g$, the distance within which the disc contains half as much magnetic flux as the black hole \citep{2012MNRAS.423L..55T}. The accretion timescale from this distance is 

\begin{equation}
t_{\rm accr} \simeq 10^4 \left(\frac{\alpha}{0.1}\right)^{-1} \left(\frac{H/R}{0.3}\right)^{-2} \left(\frac{r}{20 \, r_g}\right)^{3/2} \, r_g/c,
\end{equation}

\noindent where $\alpha$ is the dimensionless viscosity parameter and $H/R$ is the disc scale height. This order of magnitude estimate gives the upper limit for the flare recurrence timescale. Typical recurrence times for major eruptions in our fiducial simulation are $\sim 10^{3\text{--}4} \, r_g/c$ ($\simeq 5$--$50$h for Sgr A*), broadly consistent with this estimate.

These magnetic flux eruptions launch low-density tubes of magnetic flux from near the black hole, which form a rotating spiral pattern \citep{igumenshchev2008,tchekhovskoy2011}.  \autoref{fig:sim_plots_temaps} shows density-weighted, vertically averaged maps of particle density, electron temperature, plasma $\beta$ (the ratio of gas to magnetic pressure), and fluid heating rate, $Q$, at four snapshots near the $t = 8 \times 10^3 r_g/c$ eruption event in our fiducial simulation. Hot, strongly magnetized, low density plasma forms a spiral structure which rotates around the black hole at small radii of $r \lesssim 10 r_g$. Particularly strong heating (red regions in fourth row of \autoref{fig:sim_plots_temaps}) occurs near the boundary between the strongly magnetized (plasma $\beta < 0.1-1$), low density and more weakly magnetized (plasma $\beta \simeq 1-10$), higher density regions. Similar non-axisymmetric spirals are seen in all eruption events. Note that the unphysical $Q<0$ values (blue regions in fourth row of \autoref{fig:sim_plots_temaps}) occur due to unavoidable truncation error in low density, high entropy, high magnetization regions (see \S 6 of \citealt{ressler2017} and \S 3.1 of \citealt{sadowski2017} for discussions of this issue).  These regions are excluded from our emission calculations (\autoref{sec:niremission}).

\begin{figure*}
\begin{tabular}{cc}
\includegraphics[width=0.45\textwidth]{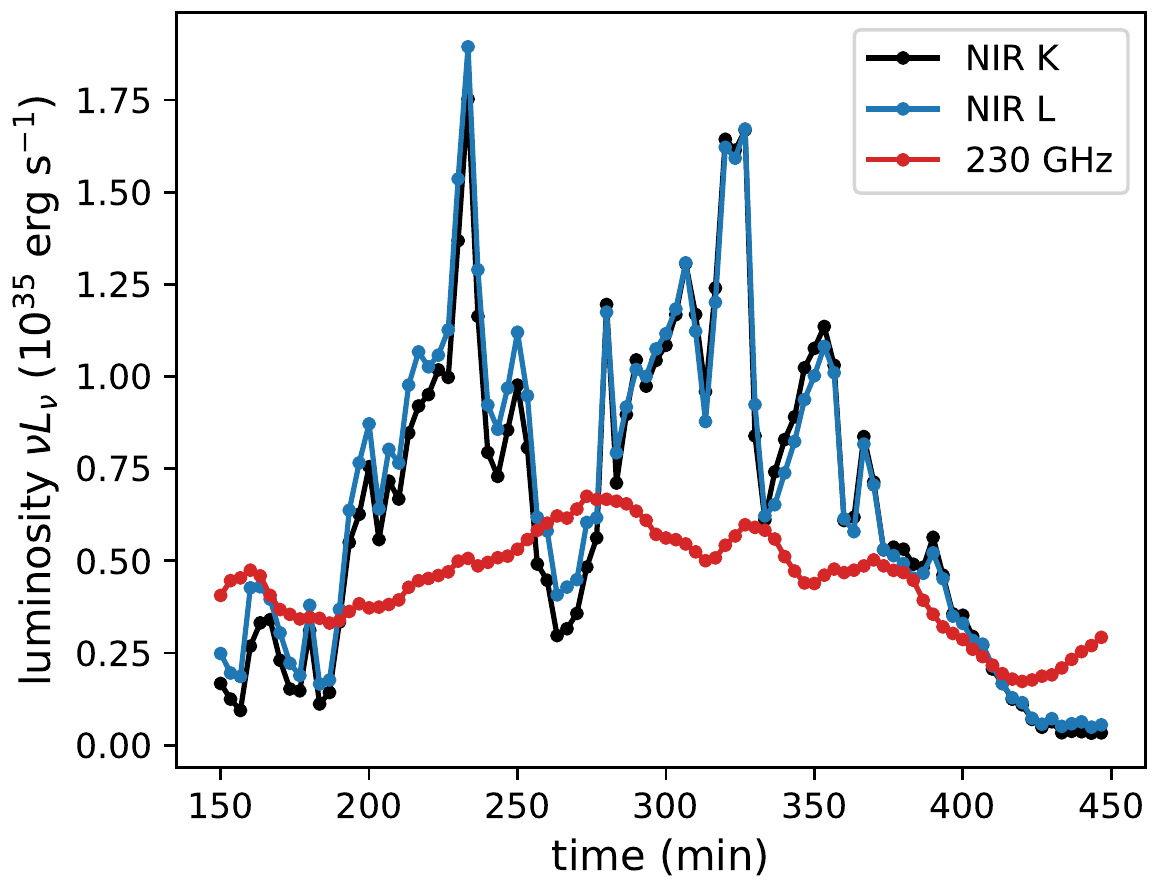} & 
\includegraphics[width=0.45\textwidth]{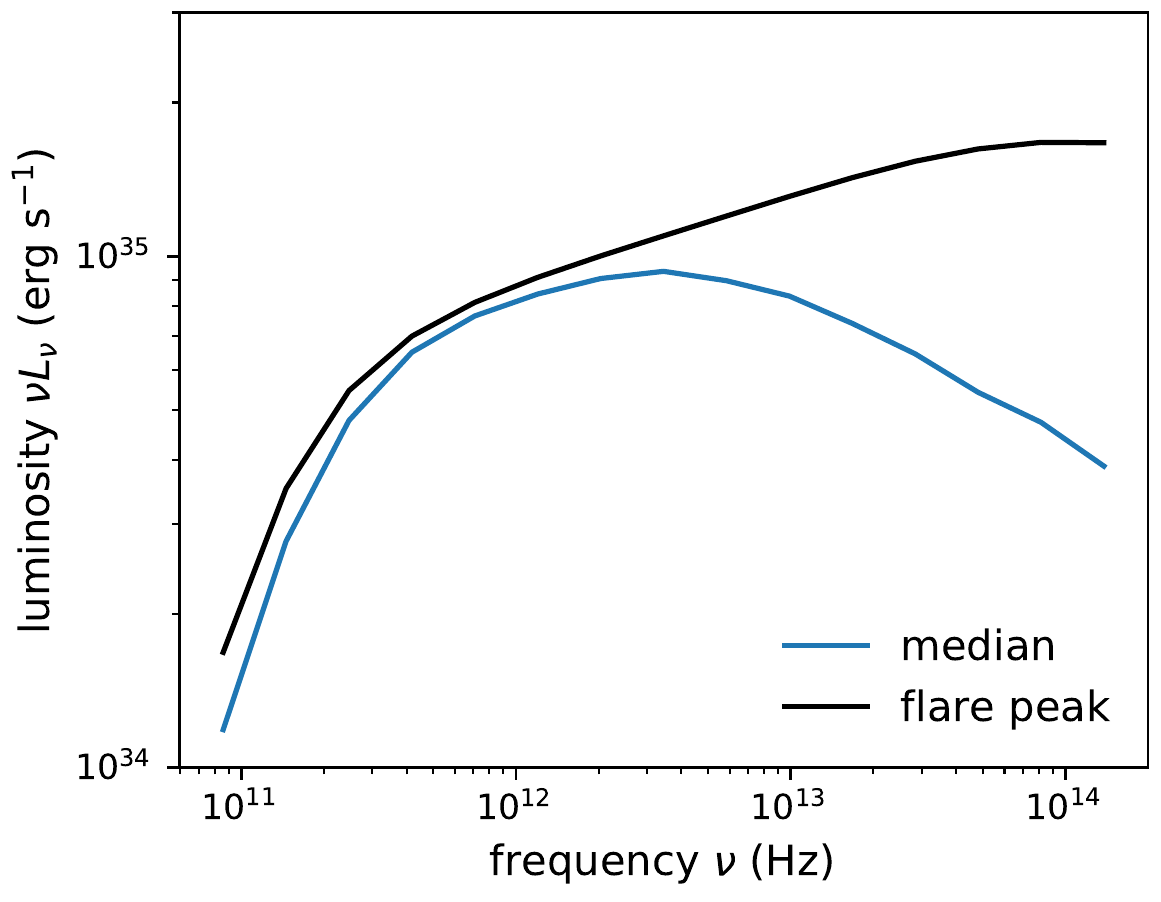}
\end{tabular}
\caption{\label{fig:nir_submm_lcurve}Left: K ($2.2$ $\mu$m) and L ($3.8$ $\mu$m) NIR and $1.3$ mm (230 GHz) light curves during the flare studied here. The NIR spectral index in $\nu L_\nu$ varies between $\simeq -1$--$0.2$. The submm light curve shows its typical factor $\simeq 50\%$ variability, without a simultaneous change  corresponding to the NIR flare. Right: median spectral energy distribution over time compared to that of the flare peak. Flaring occurs in the infrared, with little effect on the submm part of the spectrum.}
\end{figure*}

\section{Near-infrared observables}
\label{sec:niremission}

We calculate NIR observables from the time period corresponding to this same magnetic flux eruption event. We scale the simulation data to cgs units using an average mass accretion rate of $10^{-8} M_\odot$~yr$^{-1}$ \citep[chosen to match the Sgr A* submm to NIR spectrum,][]{dexter2020harmpi}. We calculate the NIR emissivity assuming a thermal energy distribution of electrons using the angle-dependent fitting function from \citet{mahadevan1996}. 
We then calculate polarized movies of synchrotron radiation using the \texttt{grtrans}\footnote{\url{https://github.com/jadexter/grtrans}} code \citep{dexter2009,dexter2016} including all relativistic effects. We fix the observer inclination angle at $i = 25^\circ$, consistent with GRAVITY results \citep{gravity2018flare,bauboeck2020} except that our fluid is rotating counterclockwise on sky. Images are calculated with a field of view of $130 \, \mu$as with $300 \times 300$ pixels. We exclude any emission from the most highly magnetized regions with the ratio of magnetic to rest mass energy density $b^2 / \rho > 1$, with both quantities in code units. We fix the mass of Sgr A* at $M = 4 \times 10^6 M_\odot$  with a distance to the Galactic center of $D = 8$ kpc \citep[e.g.,][]{ghez2008,gillessen2009,gravity2019,do2019,gravity2020prec}. These choices lead to characteristic length, time, and angular scales of $r_g \simeq 6 \times 10^{11}$ cm, $r_g/c \simeq 20$ s, and $r_g/D \simeq 5 \, \mu$as.

The black curve in \autoref{fig:sim_plots_phibh_lcurve} shows the NIR model light curve over a continuous $\simeq 300$h time interval for Sgr A*. Strong peaks in near-infrared flux density are apparent, each corresponding to a sharp decrease in magnetic flux on the event horizon. Sample snapshot images are shown in the last row of \autoref{fig:sim_plots_temaps}. The spiral morphology matches that seen in the coordinate frame simulation data of the same flare, particularly that of the fluid heating rate $Q$. The emission region rotates continuously completing one full period over this time span. The left panel of \autoref{fig:flux_astro} shows one double-peaked K band ($2.2 \, \mu$m) NIR flare corresponding to the time period of the eruption event studied above. The $x$ and $y$ centroids also show motion suggestive of rotation, with an on sky diameter of $\simeq 20-40 \, \mu$as ($4-8 \, r_g$). We show the astrometric motion on sky in the right panel. Here we use simulation data spaced by $\Delta t = 20$s and account for light travel time delays through the emission region (``slow light"). The NIR centroid rotates continuously over the $80$ minutes. \autoref{fig:qu} shows that this rotation is accompanied by a single rotation of the linear polarization angle on the same time scale, which shows up as oscillations of the fractional Stokes parameters $Q/I$ and $U/I$ \citep[e.g.,][]{marrone2006conf}.

\autoref{fig:nir_submm_lcurve} compares NIR $K$ and $L$ band light curves with that at $1.3$mm (230 GHz). The spectral index is variable during the flare but fairly flat, ranging from values of $ -1 \lesssim \alpha \lesssim 0.2$ for $\nu L_\nu \sim \nu^\alpha$. There is no simultaneous rise in submm flux during the NIR flare, but rather the emergence of a new higher energy spectral component (see median and flare spectral energy distributions in the right panel of \autoref{fig:nir_submm_lcurve}).

Where does the emission originate? The flux eruptions are associated with magnetically dominated, low density material. Synchrotron radiation scales with density, and we explicitly exclude emission from highly magnetized regions where $b^2 / \rho > 1$. The observed emission morphology tracks that of the magnetized material, but most closely with the regions of highest dissipation at the interface between magnetically dominated regions and the rest of the fluid. Typical fluid properties calculated as intensity-weighted averages are $n \simeq 10^6 \, \rm cm^{-3}$, $B \simeq 100$ G, $T_e \simeq 10^{12}$ K, $\beta \simeq 5$. The temperature is one order of magnitude higher than that of the submm emitting electrons. Still, the bulk of the NIR radiation does not originate from magnetically dominated plasma. If we instead apply a cut where $b^2 / \rho > 25$, the flux density increases by a factor $\simeq 2$. The average properties of radiating electrons all change by a simlar amount, e.g. the new emission comes from more highly magnetized material. The emission region morphology, centroid motion, and polarization behavior remain the same.

\section{Discussion}
\label{sec:discussion}

We have shown that flux eruption events associated with magnetically arrested discs produce non-axisymmetric, magnetically dominated regions which travel outwards and heat the plasma via magnetic reconnection. These events are a direct consequence of strong magnetic fields near the black hole event horizon, which become dynamically important and repel accreting gas. The existence of such a flow structure near Sgr A* may be a natural consequence of the accretion of weakly magnetized stellar winds in the central parsec \citep[][]{ressler2019,ressler2020}. 

By calculating NIR observables of Sgr A* from a fiducial 3D GRMHD simulation, we have shown that such events trigger large-amplitude NIR variability which matches many observed properties of the NIR/X-ray flares:

\begin{itemize}
    \item factor of $\simeq $10--20 increases in flux density compared to the median with durations of $\simeq$ 30--60 minutes;
    \item a recurrence timescale of several hours;
    \item a flat spectral index in the NIR, without a simultaneous submillimeter counterpart;
    \item linear polarization fractions of $\simeq $10--20$\%$;
    \item continuous rotation of the emission region accompanied by a rotation of the linear polarization angle.
\end{itemize}

\noindent These flaring events occur in a model which satisfies many other Sgr A* submm to NIR observational constraints \citep{dexter2020harmpi}. The flare recurrence time is the timescale for magnetic flux to accumulate on the black hole and saturate following a dissipation event. The flares are driven primarily by increases in electron temperature due to particle heating from magnetic reconnection. The polarization oscillation is due to strong poloidal magnetic field components in the inner MAD accretion flow \citep[][]{gravity2018flare}. While we do see rapid, large-amplitude variations in intensity in SANE simulations, those events do not show significant centroid motion or variations in the linear polarization angle.

The flares in our MAD model could thus be a promising explanation for the observed Sgr A* NIR flares. There are still some inconsistencies with the data. Compared to the observations, i) the flux distribution shows too many moderate and not enough very bright flares \citep{gravity2020fluxdist}, ii) the flare spectra might be too steep (too ``red''), iii) the astrometric and polarization periods are at the long end of the observed range, and iv) the astrometric pattern on sky is a factor of $\simeq$ 2--3 too small. 

One possible reason for these differences is that our physical model is overly simplistic. We assume a purely thermal distribution of electrons, while relativistic magnetic reconnection can produce significant non-thermal particle acceleration \citep[e.g.,][]{sironi2014,guo2015,werner2016}. Particle acceleration is commonly invoked to explain the NIR and especially X-ray flares from Sgr A* \citep[e.g.,][]{markoff2001,yuan2004,ball2016,chael2018}. The cooling time given the typical parameters of our radiating electrons is $\sim 10$ min, similar to the dynamical time close to the black hole and shorter than the flare duration. Radiative cooling may be important, particularly if higher energy non-thermal electrons contribute significantly to the observed flux. On the other hand, we do see continuous heating of the flaring electrons in our simulation (\autoref{fig:sim_plots_temaps}). We do not include non-thermal emission or Compton scattering, and so are at present unable to make predictions for the X-ray luminosity or spectra of the flaring events studied here. Using the one zone prescription of \citet{chiaberge1999} and our average NIR emission parameters above, we estimate an X-ray luminosity of $\simeq 10^{33} \, \rm erg \, \rm s^{-1}$ for synchrotron self-Compton and $\simeq 3\times 10^{34} \, \rm erg \, \rm s^{-1}$ for Compton upscattering of NIR seed photons by submm electrons. The results suggest that Compton cooling is likely sub-dominant (Compton $y \lesssim 1$). At the same time, the one zone estimate of the inverse Compton luminosity estimate is only a factor of a few below that seen in Sgr A* X-ray flares. Our brightest flare peak is a factor of $\lesssim 2$ fainter than the brightest observed NIR flare to date \citep{do2019flare}, and \citet{gutierrez2020} suggested that MAD eruptions could be energetic enough to produce this event.

There are also numerical complications inherent to modeling MADs. Strongly magnetized regions are difficult to evolve accurately in ideal GRMHD simulations such as those used here. Flux eruptions are particularly challenging in this regard, since they produce steep gradients in magnetization over a large part of the inner accretion flow. Our results qualitatively match those in previous MAD simulations \citep[e.g.,][]{tchekhovskoy2011,white2019,2020MNRAS.494.3656L}. We have also carried out otherwise identical simulations at lower resolutions of $3/4$ and $1/2$ the number of cells in each dimension. The flow structure and the time evolution of ramp up and dissipation cycles in magnetic flux are consistent in all cases (see appendix \ref{app:resolution}). Still, the robustness of the (thermo)dynamics of such events to changes in resolution,  simulation density floors, or the grid scale dissipation in ideal MHD remains uncertain. In addition, while the magnetic dissipation in the simulations here does arise in current sheets near the disc midplane (see appendix \ref{app:currents}), we do not resolve the physics of the magnetic reconnection process itself at our limited spatial resolution and using ideal MHD.

In our models, there is an average $\lesssim 10 \, \mu$as offset between the NIR emission region centroid and the position of the black hole. The offset, which depends on the chosen inclination angle (here $i = 25^\circ$), is in the direction of approaching material and results from Doppler beaming due to relativistic motion. Since our models underproduce the observed amplitude of centroid motion seen in NIR flares, we consider this a lower limit to the bias that would be induced in GRAVITY astrometric measurements in the S2 orbit in 2017 and 2018. A $\lesssim 10 \, \mu$as offset currently causes negligible bias in parameters inferred from the orbit of S2 \citep{gravity2020prec}.

We find that NIR centroid motions are larger during flares than otherwise. All flares in the fiducial simulation are associated with some degree of continuous rotation, showing apparent periods of $40$--$100$ min and completing $1/2$--$2$ rotations. The observed rotation speed is consistent with the (sub-Keplerian) orbital speed at $r \simeq 4$--$8 \, r_g$, comparable to the outer radius of the magnetically dominated structure during flares. We do not find correlations between centroid size or astrometric period and the total radiated energy or peak flux during a flare. We also see similar periods and centroid excursions in a small number of flux eruption events in shorter duration $a = 0$ and $a = 0.5$ simulations. The saturated flux values and recurrence times of eruption events appears similar in those models. \citet{ressler2020} found similar eruption events in an $a=0$ simulation with initial conditions taken from stellar wind feeding, but with longer recurrence times. According to our model, future flares should show a range of periods and astrometric sizes. In higher precision data, the centroid track would appear more complex than that of a compact region undergoing orbital motion.

Although the total submm intensity does not vary simultaneously with that of the NIR, our models do show rotations of the submm polarization angle during the flares. Similar features have been seen in submm polarimetry data \citep[e.g.,][]{marronequ}. Since the flux eruption events disrupt the inner accretion flow, we generically expect that NIR flares should be accompanied by observable signatures in resolved submm images with the Event Horizon Telescope \citep{EHTPaperII}. 

The eruption events studied here may have implications for variable emission seen in accreting black hole systems beyond Sgr A*. The flare recurrence times of weeks seen in blazars  \citep[e.g.,][]{2009ApJ...704.1689C,2011ApJ...734...43C} and of hours seen in the early light curve of the jetted tidal disruption event Swift J1644 \citep[e.g.,][]{2014MNRAS.437.2744T} both match the $10^{3-4} \, r_g/c$ eruption recurrence times found here. Much shorter timescale variability of $10-100$ ms might be expected in the light curves of both short \citep[e.g.,][]{2019MNRAS.490.4811C} and long \citep[e.g.,][]{2015MNRAS.447..327T} gamma-ray bursts. Exploring the implications for MAD flux eruptions in the non-thermal and jetted emission from a wide range of systems is a goal of future work.

\section*{Acknowledgements}

J.D. thanks M. C. Begelman, C. F. Gammie, A. Philippov, B. Ripperda, and D. Uzdensky for helpful comments. J.D. and A.J.-R. were supported in part by a Sofja Kovalevskaja award from the Alexander von Humboldt foundation, by a CONACyT/DAAD grant (57265507), and by NASA Astrophysics Theory Program Grant 80NSSC20K0527. S.M.R. was supported by the Gordon and Betty Moore Foundation through Grant GBMF7392 and also in part by the National Science Foundation under Grant No. NSF PHY--1748958. AT was supported by the National Science Foundation AAG grants 1815304 and 1911080. The calculations presented here were carried out on the MPG supercomputers Hydra and Cobra hosted at MPCDF, and using resources supported by the NASA High-End Computing (HEC) Program through the NASA Advanced Supercomputing (NAS) Division at Ames Research Center.

\section*{Data Availability}
Simulated images and averaged simulation data products used here will be shared on reasonable request to the corresponding author.

\bibliographystyle{mnras}

\appendix

\section{Impact of grid resolution on magnetic flux evolution and flow structure}
\label{app:resolution}

\begin{figure*}
\begin{tabular}{cc}
\includegraphics[width=0.43\textwidth]{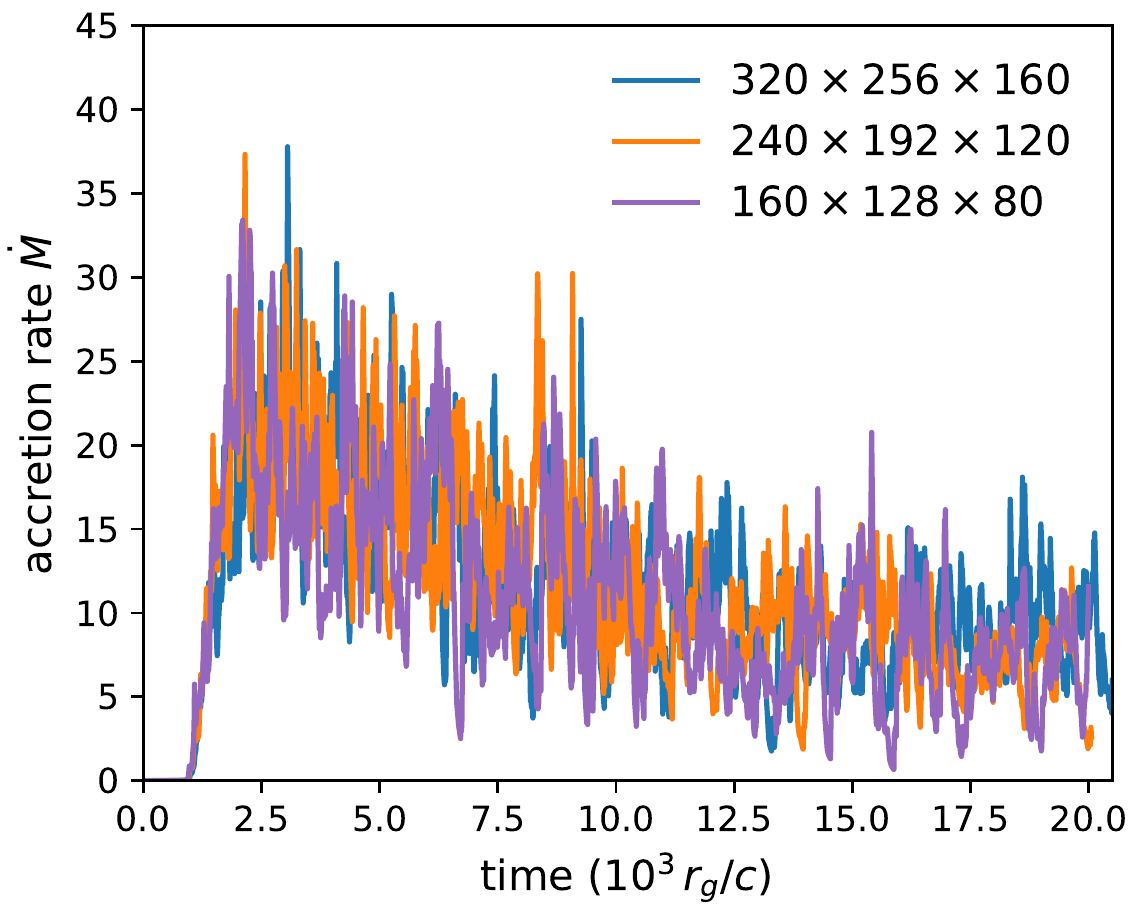}&
\includegraphics[width=0.43\textwidth]{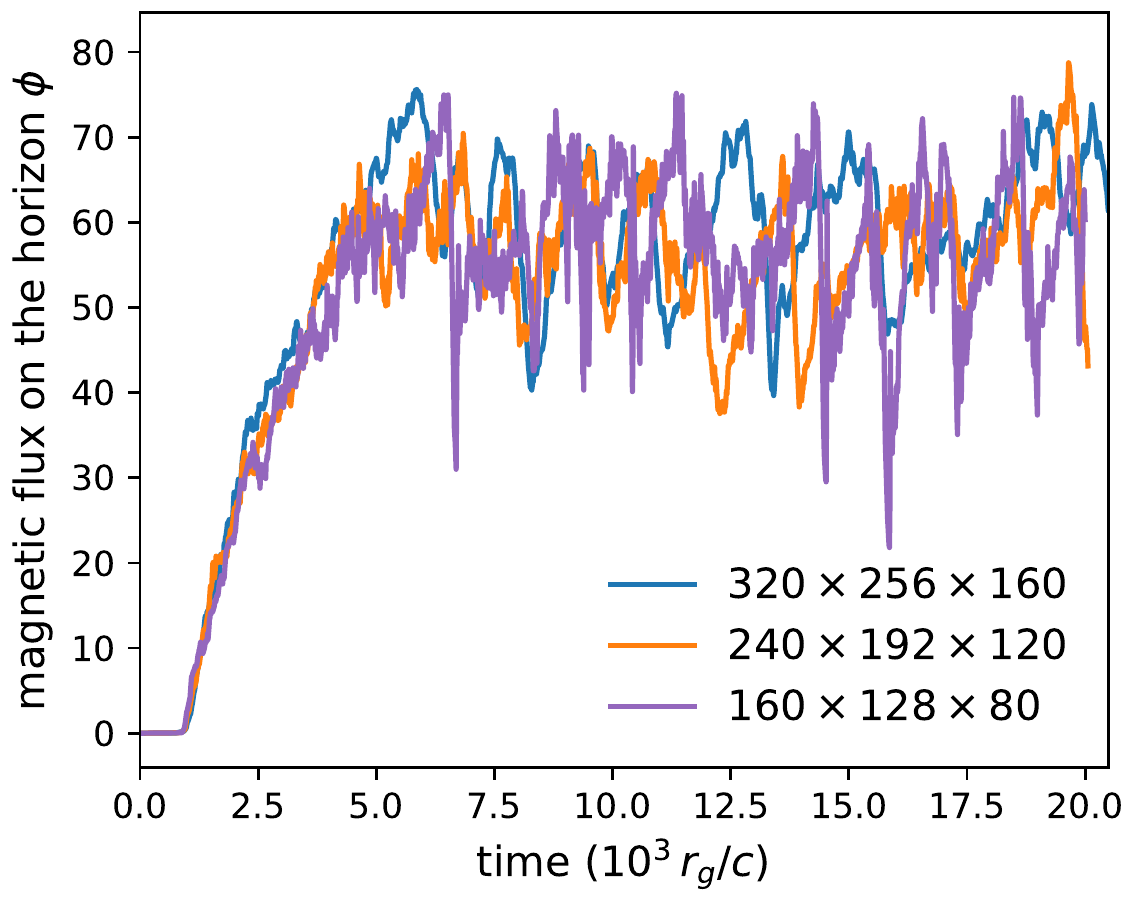}
\end{tabular}
\caption{\label{fig:madres}Mass accretion rate and dimensionless flux on the horizon for the fiducial and two lower resolution simulations, with all other parameters held constant. Both simulations show qualitatively the same evolution. The $240 \times 192 \times 120$ model also shows quantitatively similar time evolution, including e.g. the timing of flaring events and the amount of dissipated magnetic flux.}
\end{figure*}

\begin{figure*}
\begin{tabular}{cc}
\includegraphics[width=0.43\textwidth]{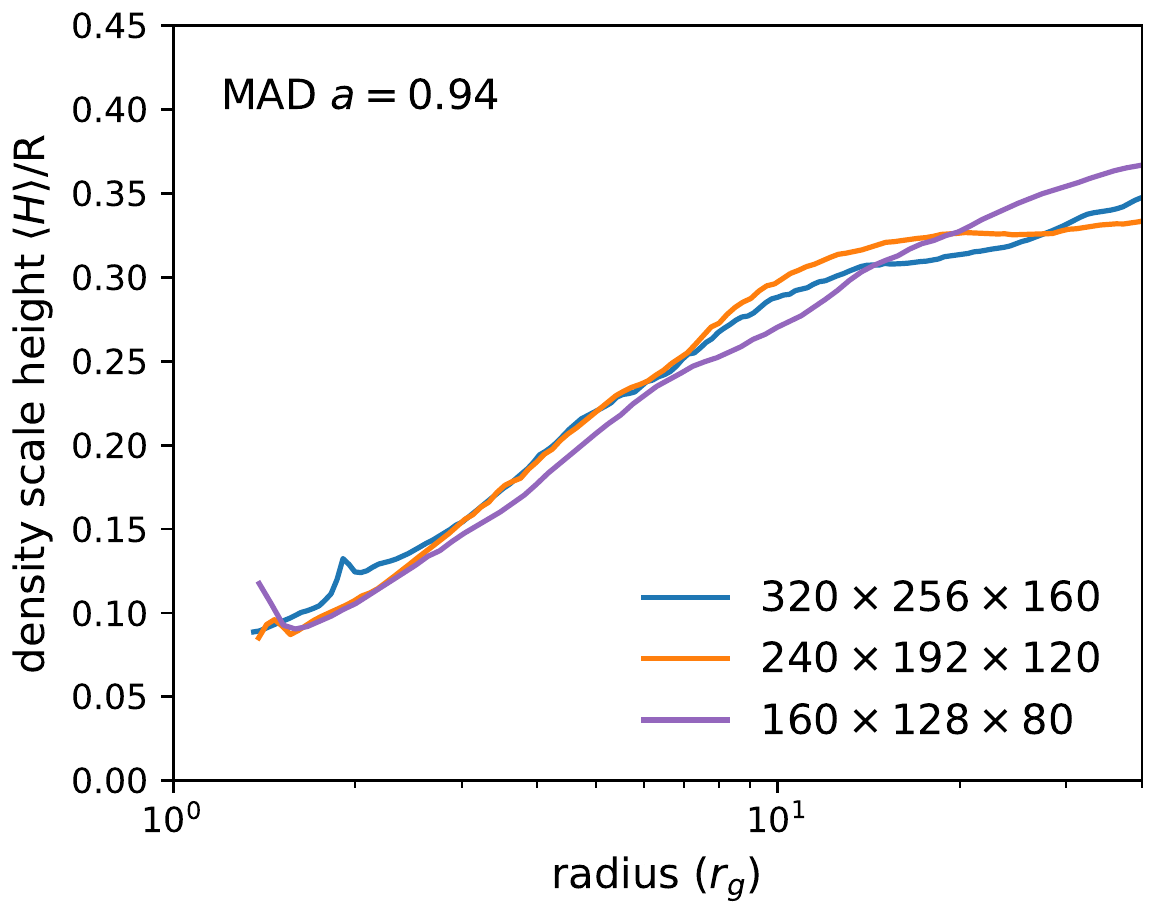} &
\includegraphics[width=0.43\textwidth]{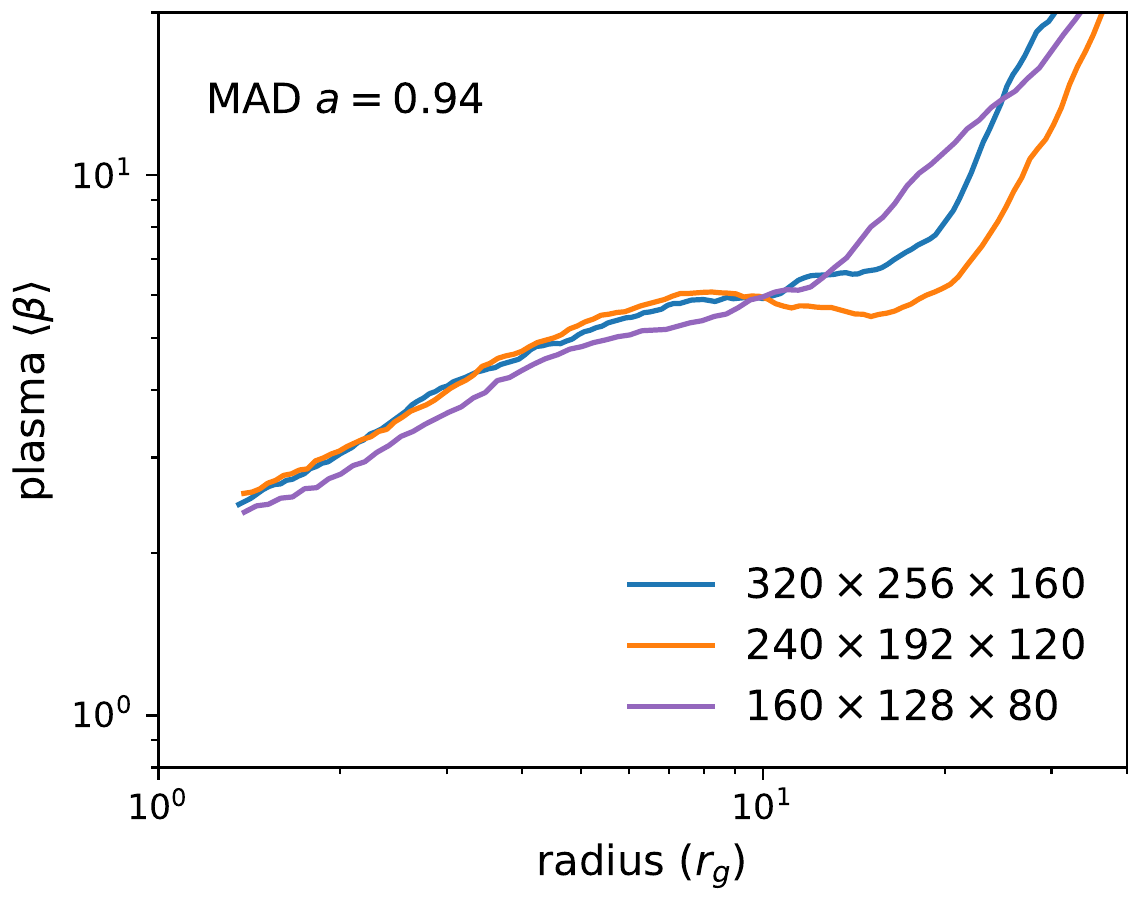}
\end{tabular}
\caption{\label{fig:madres_r}Shell-averaged scale height (left) and plasma $\beta$ (right) profiles averaged from $t = 8000$--$10000 \, r_g/c$ for the three models. All show similar time-averaged radial profiles, particularly at small radii $\lesssim 10 \, r_g$.}
\end{figure*}

To study convergence, we have run additional simulations at $3/4$ and $1/2$ the number of cells in each grid dimension as our fiducial simulation. All other parameters are kept the same, and the lower resolution simulations are run for a time $t = 2\times10^4 \, r_g/c$. \autoref{fig:madres} shows the time evolution of the mass accretion rate through the event horizon $\dot{M}$ and the dimensionless magnetic flux accumulated on the black hole $\phi$ for the fiducial and lower resolution simulations. We see qualitatively similar time evolution in all cases, including ramp up and dissipation cycles of magnetic flux on similar timescales and with similar amplitudes. \autoref{fig:madres_r} shows density-weighted shell-averaged radial profiles of the disc scale height $\langle H \rangle / R$ and plasma $\beta$. Inside of $r \lesssim 20 r_g$, the results are nearly identical for all resolutions. Many individual eruptions occur at similar times and look similar between the full and $3/4$ resolution cases in terms of vertically integrated maps like those shown in \autoref{fig:sim_plots_temaps}. In particular, they show spiral structures of erupting, highly magnetized plasma from near the black hole. We conclude that the results presented here would not change if we were to use a somewhat smaller grid resolution for our fiducial simulation.

\section{Magnetic field structure}
\label{app:currents}

\begin{figure*}
\begin{tabular}{ccc}
\includegraphics[width=0.27\textwidth]{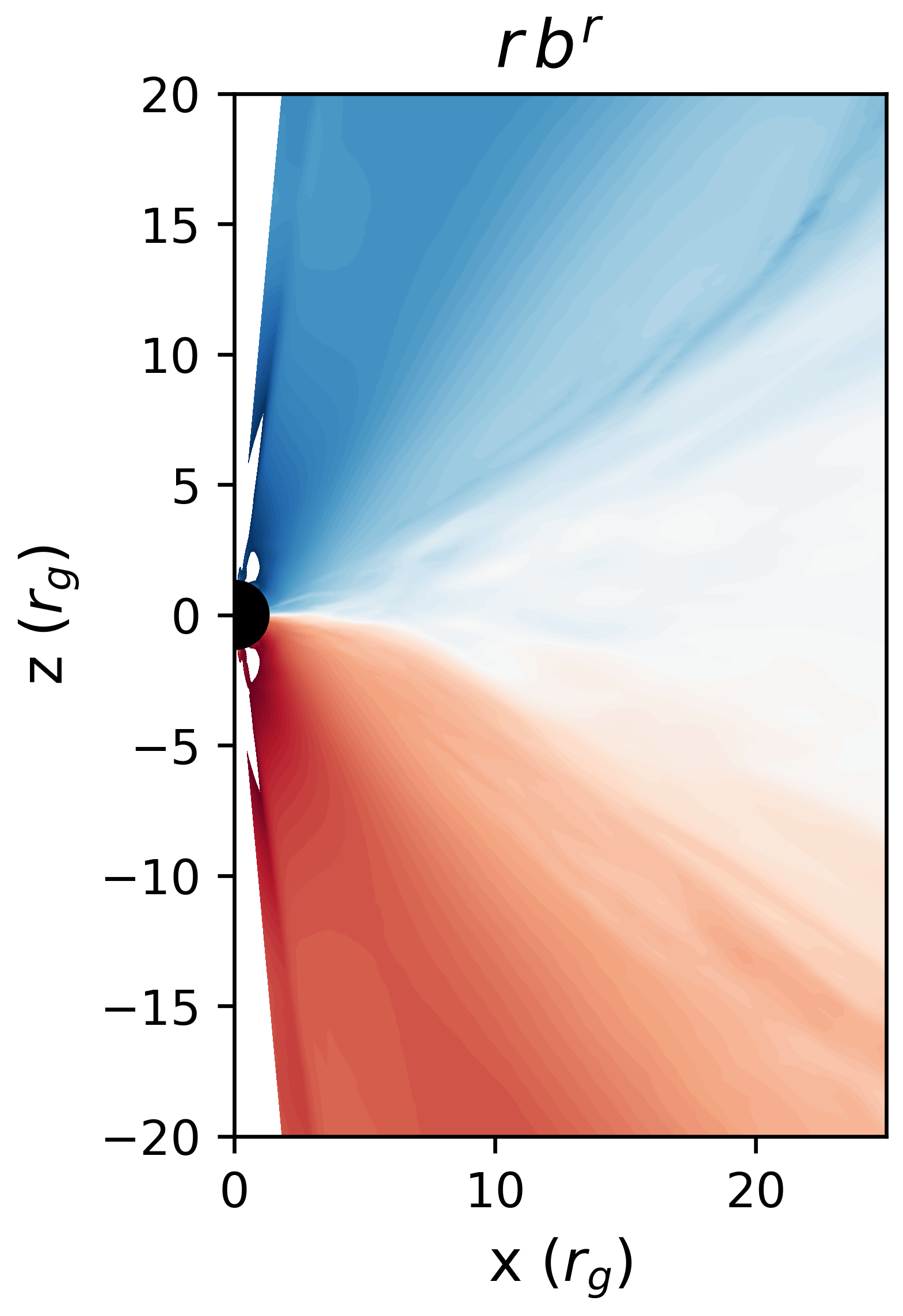}&
\includegraphics[width=0.27\textwidth]{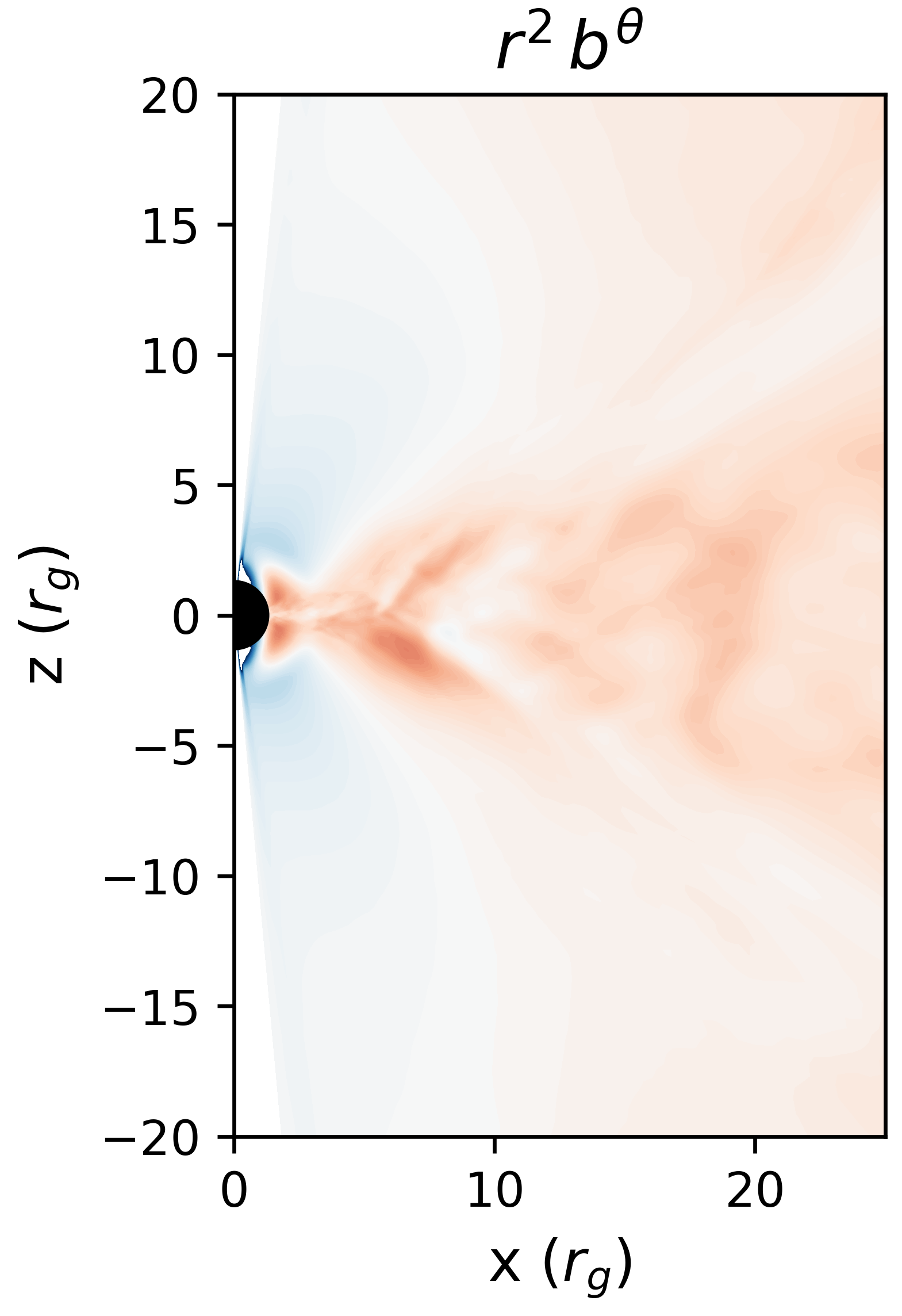}&
\includegraphics[width=0.35\textwidth]{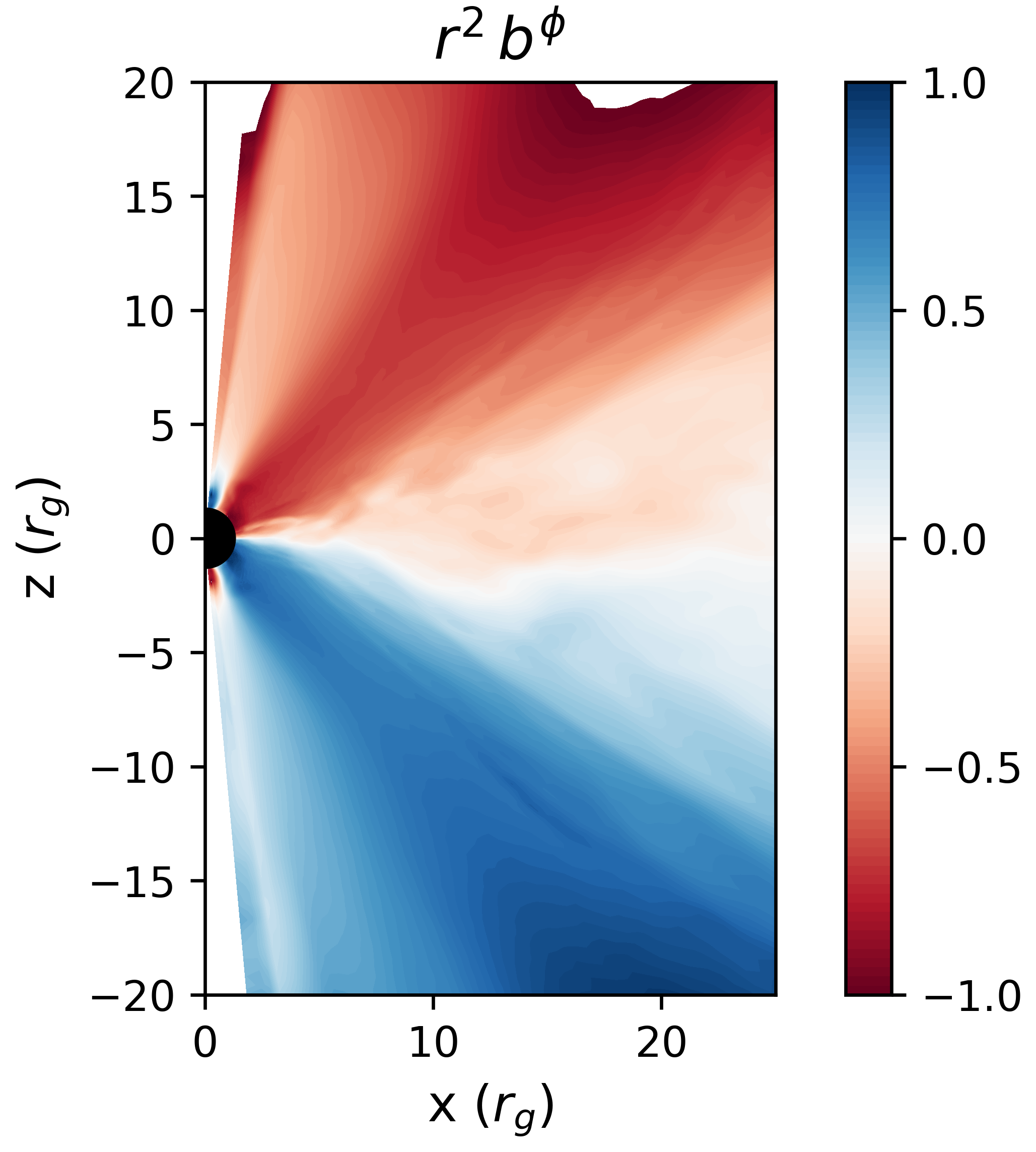}
\end{tabular}
\caption{\label{fig:madb}Azimuthally averaged snapshots of coordinate frame field components at $t = 7800 \, r_g/c$. Each component is weighted by $r$ or $r^2$ and normalized to its maximum value. Sign reversals are present in both the $r$ and $\phi$ components, indicating the presence of current sheets.}
\end{figure*}

\autoref{fig:madb} shows sample snapshots of the magnetic field structure in our simulation, at a time of $t = 7800 \, r_g/c$ during the flare studied here. Sign reversals are present in the $r$ and $\phi$ components, indicating the presence of a midplane current sheet. Future studies at higher resolution and/or with explicit resistivity \citep[e.g.,][]{nathanail2020,ripperda2020} may be able to determine which components are responsible for the observed magnetic reconnection and measure key parameters such as the relative guide field strength.

\bsp	\label{lastpage}
\end{document}